\documentclass[useAMS,usenatbib]{mn2e}
\pdfoutput=1 
\usepackage{hyperref}
\usepackage{amsmath}
\usepackage{graphicx}
\usepackage{txfonts}
\usepackage{natbib}

\newcommand{\gaia}{{\it Gaia }}
\newcommand{\g}{$G$ }
\newcommand{\bp}{$G_{\rm{BP}}$ }
\newcommand{\rp}{$G_{\rm{RP}}$ }
\newcommand{\bpgrp}{$G_{\rm{BP}}$, $G$, \rm{and} $G_{\rm{RP}}$ }
\newcommand{\teff}{T_{\rm{eff}}}

\newcommand{\logg}{\log\,g}
\newcommand{\feh}{\rm{[Fe/H]}}
\newcommand{\aFe}{[\alpha/\rm{Fe}]}

\newcommand{\flux}{\rm{erg\,s^{-1}\,cm^{-2}}}

\newcommand\lta{\mathrel{\hbox{\raise 0.6 ex \hbox{$<$}\kern
                   -1.8 ex\lower .5 ex\hbox{$\sim$}}}}
\newcommand\gta{\mathrel{\hbox{\raise 0.6 ex \hbox{$>$}\kern
                   -1.7 ex\lower .5 ex\hbox{$\sim$}}}}
\newcommand{\RN}[1]{{\small\textup{\uppercase\expandafter{\romannumeral#1}}}}

\title[On the use of {\it Gaia} photometry]{On the use of {\it Gaia}
  magnitudes and new tables of bolometric corrections}

\author[Casagrande \& VandenBerg]{\parbox{18cm}{
    L.~Casagrande$^{1,2}$\thanks{Email:luca.casagrande@anu.edu.au},
    Don A.~VandenBerg$^{3}$}\\
  \phantom{,}$^1$ Research School of Astronomy and Astrophysics, Mount Stromlo 
  Observatory, The Australian National University, ACT 2611, Australia\\
  $^2$ ARC Centre of Excellence for All Sky Astrophysics in 3 Dimensions
  (ASTRO 3D)\\
  $^3$ Department of Physics \& Astronomy, University of Victoria, P.O.~Box 
  1700 STN CSC, Victoria, BC, V8W 2Y2, Canada}

\begin{document}

\date{Received; accepted}

\maketitle

\begin{abstract}
  The availability of reliable bolometric corrections and reddening estimates,
  rather than the quality of parallaxes will be one of the main limiting
  factors in determining the  luminosities of a large fraction of \gaia stars.
  With this goal in mind, we provide \gaia \bpgrp synthetic photometry for the
  entire MARCS grid, and test the performance of our synthetic colours and
  bolometric corrections against space-borne absolute spectrophotometry. We
  find indication of a magnitude-dependent offset in \gaia DR2 \g magnitudes,
  which must be taken into account in high accuracy investigations. 
  Our interpolation routines are easily used to derive bolometric corrections at
  desired stellar parameters, and to explore the dependence of \gaia photometry
  on $\teff$, $\logg$, $\feh$, $\aFe$ and $E(B-V)$. \gaia colours for the Sun
  and Vega, and $\teff$-dependent extinction coefficients, are also provided. 
\end{abstract}


\section{Introduction}

\gaia DR2 includes photometry in the \bpgrp bands for approximately 1.5
billion sources. Its exquisite quality will define the new standard in the
years to come, and have a tremendous impact on various areas of astronomy. The
first goal of this letter is to lay out the formalism to generate \gaia
magnitudes from stellar fluxes, using the official \gaia zero-points
and transmission curves \citep{evans}. In doing so, we discuss in some detail
the effect of different zero-points, and search for an independent validation
of the results using space-based spectrophotometry. 

One of the important prerequisites for stellar studies is the availability
of \gaia colour transformations and bolometric corrections (BCs) to transpose
theoretical stellar models onto the observational plane, and to estimate
photometric stellar parameters, including luminosities. In
\citet[hereafter Paper I and II]{cv14,cv18} we
have initiated an effort to provide reliable synthetic colours and
BCs from the MARCS library of theoretical stellar fluxes \citep{g08} for
different combinations of $\teff$, $\logg$, $\feh$, $\aFe$ and $E(B-V)$. Here,
we extend that work to include the \gaia system. We also provide extinction
coefficients and colours for Vega and the Sun, both being important
calibration points for a wide range of stellar, Galactic, and extragalactic
astronomy. 
\begin{figure*}
\begin{center}
\includegraphics[width=1\textwidth]{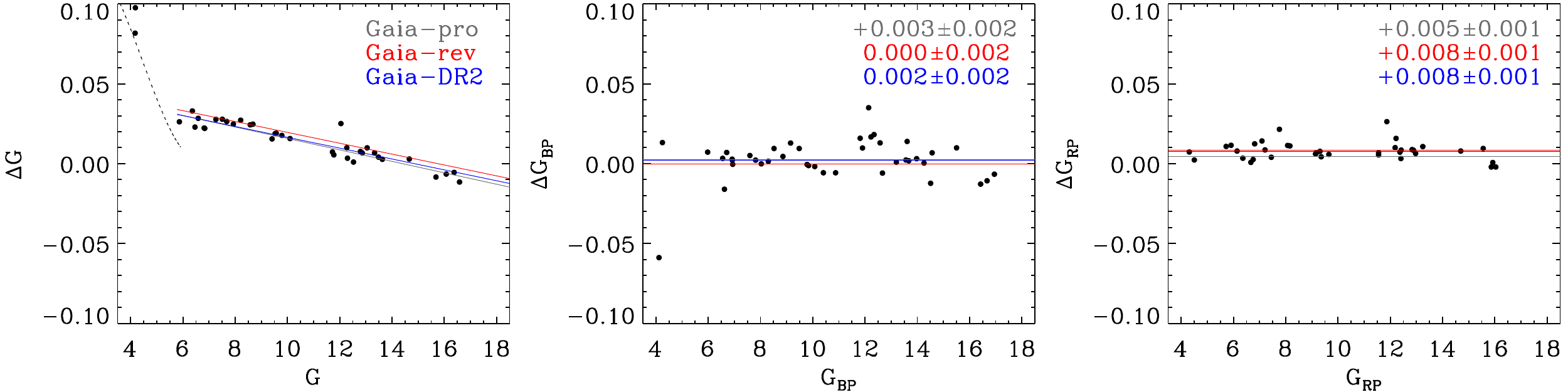}
\caption{Residuals between synthetic photometry computed with the CALSPEC
  library and the corresponding Gaia magnitudes. Only the comparisons with the
  {\tt Gaia-DR2} synthetic magnitudes are shown, but nearly indistinguishable
  trends are obtained using {\tt Gaia-pro} and {\tt Gaia-rev} (see Section
  \ref{sec:gaia} for nomenclature). Median residuals and standard deviations
  of the mean are reported for all cases. The departure at $G\sim4$
  (left-hand panel), which is not included as part of our fit, is likely due
  to the saturation of bright sources in {\it Gaia}. The dotted line in the 
  left-hand panel is the correction at bright magnitudes from
  \citet[][their Eq.~(B1)]{evans}.}\label{fig:zp}
\end{center}
\end{figure*}

\section{The \gaia system}\label{sec:gaia}

The precision of \gaia photometry is better than that of any other
currently available large catalogs of photometric standards; hence its
calibration is achieved via an internal, self-calibrating method
\citep{carrasco}. This robust, internal photometric system is then tied to the
Vega system by means of an external calibration process that uses a set of
well observed spectro-photometric standard stars \citep{pancino,altavilla}.
Observationally, a \gaia magnitude is defined as:
\begin{equation}\label{obsm}
m_\zeta = -2.5\log \bar{I_\zeta} + ZP_\zeta
\end{equation}
where $\bar{I_\zeta}$ is the weighted mean flux in a given band $\zeta$
(i.e., $G_{\rm{BP}}$, $G$ or $G_{\rm{RP}}$), and $ZP_\zeta$ is the zero-point to
pass from instrumental to observed magnitudes \citep{carrasco}. Zero-points are
provided to standardize \gaia observations to the {\tt Vega}
($ZP_{\zeta,{\tt VEGA}}$) or {\tt AB} ($ZP_{\zeta,{\tt AB}}$) systems. 
The weighted mean flux measured by \gaia for a spectrum $f_{\lambda}$ can be
calculated from:
\begin{equation}\label{eq:2}
\bar{I_\zeta} = \frac{P_A}{10^{9}hc} \int \lambda f_{\lambda} T_\zeta \rm{d}\lambda
\end{equation}
where $P_A$ is the telescope pupil area, $T_\zeta$ the bandpass, $h$ the Planck
constant, and $c$ the speed of light in vacuum \citep[see][for the units of
  measure in each term]{evans}. While $T_\zeta$, $ZP_{\zeta,{\tt VEGA}}$ and
$ZP_{\zeta,{\tt AB}}$ are the quantities used to process and publish the \gaia DR2
photometry, \cite{evans} also provide a revised set of transmission curves and
zero-points ($T_\zeta^R$, $ZP_{\zeta,{\tt VEGA}}^{\,R}$ and
$ZP_{\zeta,{\tt AB}}^{\,R}$). Here, we have generated synthetic photometry using
both sets, and call them ``processed'' ({\tt Gaia-pro}) and ``revised''
({\tt Gaia-rev}). Although the revised transmission curves and zero-points
provide a better characterization of the satellite system, DR2 magnitudes were
not derived using them. To account for this inconsistency, the published DR2
{\tt Vega} magnitudes should be shifted by
$-ZP_{\zeta,{\tt VEGA}}+ZP_{\zeta,{\tt VEGA}}^{\,R}$ \citep{brown}. Since many users
might overlook this minor correction (at the mmag level), we supply a third
set of synthetic magnitudes that take it into account, by using
$ZP_{\zeta,{\tt VEGA}}$ in Eq.~(\ref{obsm}) and $T_\zeta^R$ in
Eq.~(\ref{eq:2}). We call these magnitudes {\tt Gaia-DR2} in our
interpolation routines, and they should be preferred when comparing predicted
colors with published DR2 photometry {\it as is}.

While we adopt the formalism of Eq.~(\ref{obsm}) and (\ref{eq:2}), we remark
that from the definition of {\tt AB} magnitudes (e.g., Paper I, where
$m_{\zeta,{\tt AB}}$ or $m_{\zeta,{\tt AB}}^{\,R}$ indicates whether
$T_\zeta$ or $T_\zeta^R$ are used to compute {\tt AB} magnitudes), an alternative
formulation to generate synthetic \gaia magnitudes in the {\tt Vega} system
would be $m_{\zeta,{\tt AB}} + ZP_{\zeta,{\tt VEGA}} - ZP_{\zeta,{\tt AB}}$
({\tt Gaia-pro}), $m_{\zeta,{\tt AB}}^{\,R} + ZP_{\zeta,{\tt VEGA}}^{\,R} - ZP_{\zeta,{\tt AB}}^{\,R}$ ({\tt Gaia-rev}) and $m_{\zeta,{\tt AB}}^{\,R} + ZP_{\zeta,{\tt VEGA}} - ZP_{\zeta,{\tt AB}}^{\,R}$ ({\tt Gaia-DR2}). These hold true if the \gaia
zero-points in Eq.~(\ref{obsm}) provide exact standardization to the {\tt AB}
system. We verified that the magnitudes obtained with this alternative
formulation vs. Eq.~(\ref{obsm}) and (\ref{eq:2}) are identical to $<1$\,mmag
for {\tt Gaia-pro}, and to 3\,mmag for {\tt Gaia-rev} and {\tt Gaia-DR2}.
Importantly, we note that, in no instance, have we used Gaia DR1 data, nor the
pre-launch filter curves \citep{jordi10} anywhere in this paper.
\subsection{Check on zero-points}\label{sec:zp}

The CALSPEC\footnote{\href{http://www.stsci.edu/hst/observatory/crds/calspec.html}{http://www.stsci.edu/hst/observatory/crds/calspec.html}} library contains
composite stellar spectra that are flux standards in the {\it HST} system. The
latter is based on three hot, pure hydrogen white dwarf standards normalised to
the absolute flux of Vega at 5556 \AA. The absolute flux calibrations of CALSPEC
stars are regularly updated and improved, arguably providing the best
spectrophotometry set available to date, with a flux accuracy at the (few)
percent level \citep{bohlin14}. In particular, the highest quality measurements
in CALSPEC are obtained by the STIS ($0.17-1.01\mu$m) and NICMOS
($1.01-2.49\mu$m) instruments on board the {\it HST}.

By replacing $f_\lambda$ in Eq.~(\ref{eq:2}) with CALSPEC fluxes, it is thus
possible to compute the expected \bpgrp magnitudes for these stars, to compare
with those reported in the \gaia catalogue. For this exercise, we use only
CALSPEC stars having STIS observations (i.e. covering the \gaia bandpasses).
Further, we remove stars labelled as variable in CALSPEC, and retain only \gaia
magnitudes with the designation {\tt duplicated\_source=0},
{\tt phot\_proc\_mode=0} (i.e. ``Gold'' sources, see \citealt{riello18}).
We also remove a handful of stars with flux excess factors (a measure of the
inconsistency between \bpgrp bands typically arising from binarity, crowdening
and incomplete background modelling) that are significantly higher than those
of the rest of the sample ({\tt phot\_bp\_rp\_excess\_factor<1.3}). The
resultant comparison is
shown in Figure \ref{fig:zp}. The differences between the computed and observed
\bp and \rp magnitudes are only a few mmag. However, \g magnitudes show a clear
magnitude-dependent trend, which in fact is qualitatively in agreement with
those shown in the left-hand panels of figures (13) and (24) by \cite{evans}.
After taking into account the errors in the synthetic magnitudes from CALSPEC
flux uncertainties and \gaia measurements\footnote{In all instances, flux
uncertainties from CALSPEC and \gaia are small enough that the skewness of
mapping fluxes into magnitudes has no impact, but see Paper I, Appendix B for
a discussion of this effect.}, the significance of this slope is close to
$5\sigma$. 
No trend is observed as
function of colour. A constant offset between CALSPEC and \gaia synthetic
magnitudes would indicate a difference in zero-points or absolute calibration,
simply confirming intrinsic limitations on the current absolute flux scale
(which linchpin on Vega's flux at 5555 \AA~for CALSPEC, and 5500 \AA~for
{\it Gaia}). A drift of the CALSPEC absolute flux scale at fainter magnitudes
would appear in all filters: the fact no trend is seen for \bp nor \rp
magnitudes likely indicates that the cause of the problem stems from \gaia \g
magnitudes instead. The sign of this trend implies that \gaia \g magnitudes are
brighter than synthetic CALSPEC photometry for $G\lesssim14$, although for a few
stars this occurs at $G\sim12$. Understanding the origin of this deviation is
outside the scope of this letter. Here we provide a simple fit to place
\gaia \g magnitudes onto the same CALSPEC scale as for \bp and \rp magnitudes:
\begin{equation}\label{eq:correct}
G^{\rm{corr}} = 0.0505 + 0.9966\,G,
\end{equation}
which applies over the range $6\lesssim G \lesssim 16.5$ mag. While brighter
\g magnitudes in \gaia are affected by saturation (the trend found by
\citealt{evans} using {\it Tycho2} and {\it Hipparcos} photometry
is also seen by us, see Figure \ref{fig:zp}), it remains to be seen whether
the offset that we find extends to magnitudes fainter than $16.5$.
\begin{figure*}
\begin{center}
\includegraphics[width=1\textwidth]{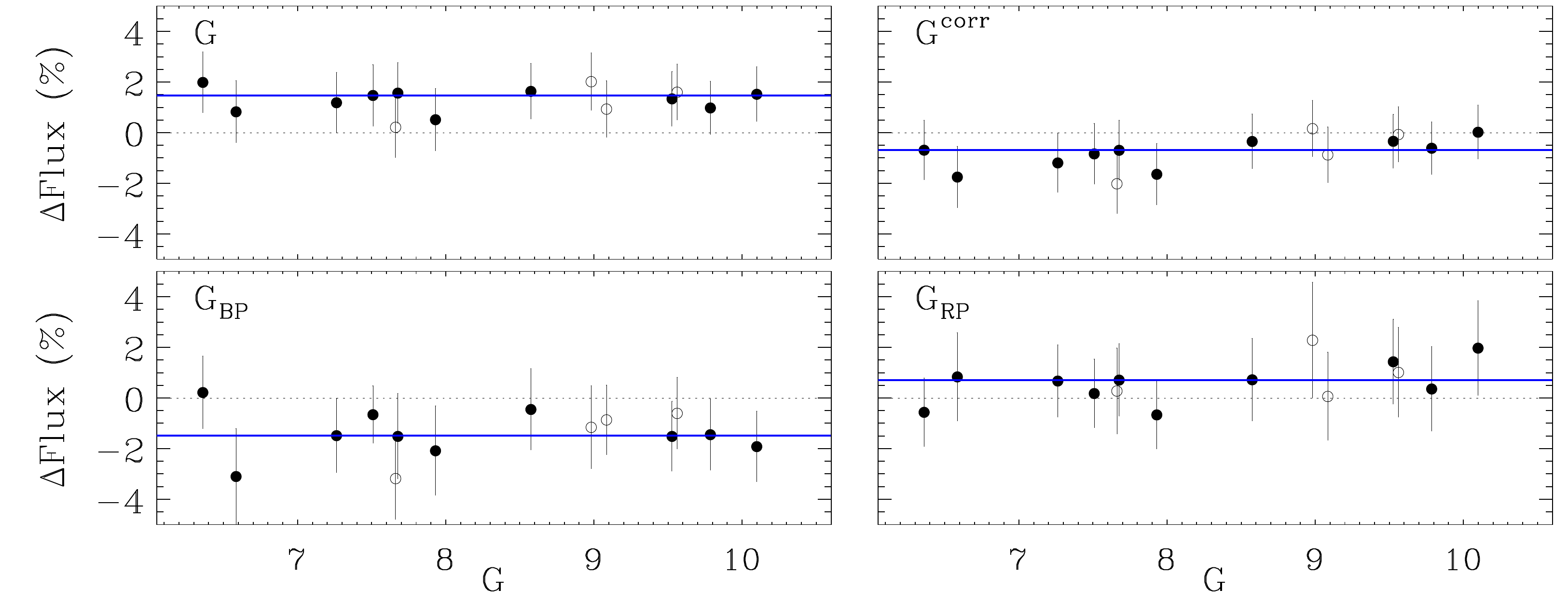}
\caption{Percentage difference between bolometric fluxes from CALSPEC
  photometry, and those recovered from our BCs (bands indicated in the top left
  corner of each panel). The stars and parameters that were adopted in our
  interpolation routines are the same as in Table 2 of Paper II. Filled circles
  are stars satisfying quality requirements listed in Section \ref{sec:zp}.
  Open circles are stars with {\tt duplicated\_source=1} (which may indicate
  observational, cross-matching or processing problems, or stellar
  multiplicity, and probable astrometric or photometric problems in all cases).
  Errors bars are obtained assuming a fixed 1 percent uncertainty in CALSPEC
  fluxes, and MonteCarlo simulations taking into account the quoted
  uncertainties in both the input stellar parameters and observed photometry
  for each target. Continuous blue lines indicate median offsets, dotted lines
  centred at zero are used to guide the eye. BCs from the
  {\tt Gaia-DR2} set are used in all instances, although nearly identical
  results are obtained using the {\tt Gaia-pro} and {\tt Gaia-rev}
  sets.}\label{fig:fluxes} 
\end{center}
\end{figure*}

\section{On bolometric corrections and other uncertainties on stellar
  luminosities}

We refer to Paper I and II for a description of the MARCS grid, our
interpolation routines, and examples of their use for different input reddenings
(in all cases we have adopted the \citealt{ccm89} parametrization of the
extinction law). We also emphasize once more the importance of paying attention
to the zero-point of the bolometric magnitude scale, which is arbitrary, but
once chosen, must be abided. In our grid, there is no ambiguity in the
zero-point of the BCs, which is instead an unnecessary 
source of biases when matching a synthetic
grid to heterogeneous observations \citep{andrae}. To derive BCs from our grid
requires the prior knowledge of stellar parameters, which often might not be a
trivial task. Our scripts easily allow one to test the effects on BCs of varying
the input stellar parameters. Projecting BCs as function of $\teff$ would also
be affected by the distribution of stellar parameters underlying the grid. In
nearly all circumstances, this distribution would be different from that of
the sample used for a given investigation. 

In Paper I and II we carried out extensive tests of the MARCS synthetic colours
and BCs against observations, concluding that observed broad-band colours are
overall well reproduced in the range encompassed by the \gaia bandpasses, with
the performance downgrading towards the blue and ultraviolet spectral regions. 
For the sake of this letter, we want to know how well bolometric fluxes can be
recovered\footnote{Bolometric flux ($\flux$) implies the flux across the entire
  spectrum that an observer (us) would measure from a star at distance $d$. On
  the other hand, luminosity ($\rm{erg\,s^{-1}}$) refers to the intrinsic
  energy output of a star, i.e., $4\,\pi\,d^2$ times the bolometric flux.} from
\gaia photometry. In fact, \gaia parallaxes deliver exquisite absolute
magnitudes
for a large fraction of stars \citep{babu}. However, when comparing them with
stellar models, one of the main limiting factors stems from the quality of the
BCs.  Here we extend the comparison of Table 2 in Paper II (which is limited by
the availability of reliable stellar parameters to F and G dwarfs at various
metallicities) to include \gaia
\bpgrp magnitudes. Our goal is to assess how well our BCs recover the
bolometric fluxes measured from the CALSPEC library. This is shown in Figure
\ref{fig:fluxes}. The first thing to notice is the offset, as well as scatter
associated with the BCs in \bp.
While the synthetic photometry presented in
Figure \ref{fig:zp} only relies on the observed spectrophotometry and how
well a bandpass is standardized, the quality of the comparison in
Figure \ref{fig:fluxes} also depends on the MARCS models, as well as the
input stellar parameters that were adopted when our tables of BCs were
interpolated. As already mentioned, the performance of the MARCS models
downgrades toward the blue, and in this spectral region stellar fluxes have
a stronger dependence on stellar parameters. The comparison is better in \rp
band (bottom right) as well as in the \g band (top panels). In the
latter case, applying Eq.~(\ref{eq:correct}) to correct the \gaia photometry
improves the agreement, but it does not yield to a perfect match (for the same
reasons that were just discussed). In all cases, the offset and scatter
typically vary between 1 and 2 percent, which we regard as the uncertainty of
our BCs (0.01--0.02 mag for the F and G dwarfs tested here).  

\begin{figure*}
\begin{center}
\includegraphics[width=1\textwidth]{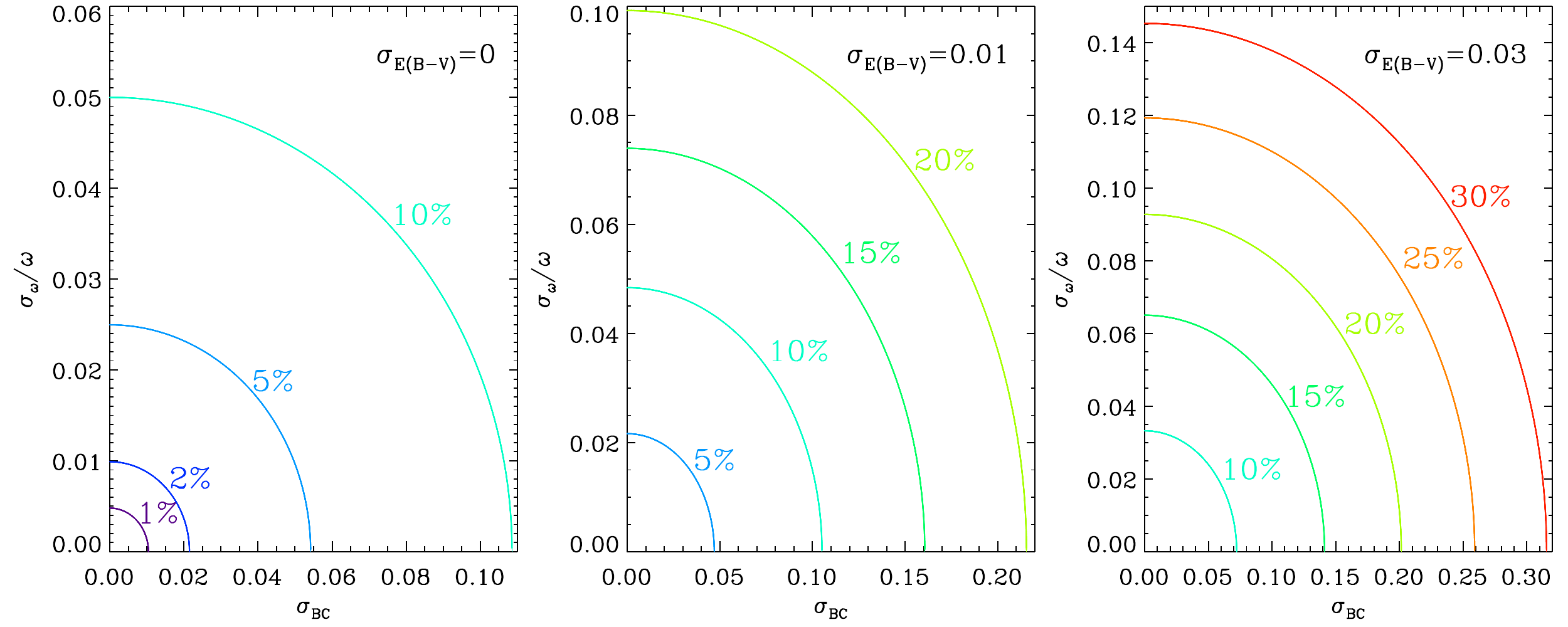}
\caption{{\it Left-hand panel:} correlation between the uncertainties in parallax
  ($\sigma_\omega/\omega$) and BCs ($\sigma_{\rm{BC}}$) for a target precision
  in luminosity (indicated by different curves). Magnitude uncertainties are
  fixed at 3\,mmag. {\it Middle and right-hand panels:} same as the left-hand
  panel, but assuming a reddening uncertainty of $0.01$ and $0.03$ mag,
  respectively. An extinction coefficient of $2.7$ is adopted (appropriate for
  the $G$ band, and in between those for \bp and \rp). The theoretical lower
  limit on the luminosity error is set by \gaia magnitudes in the left-hand
  panel ($\sim0.3$\%), and reddening in the central ($\sim2.5$\%) and right-hand
  ($\sim7.5$\%) panels.}\label{fig:errors} 
\end{center}
\end{figure*}

To summarize, we now estimate the fractional contribution of different
uncertainties
to stellar luminosities. Assuming systematic errors in magnitudes are under
control (or corrected for, as we discussed), the
precision of \gaia magnitudes $\sigma_\zeta$ is typically
at the mmag level (although larger for sources that are very bright, in crowded
regions, or very faint) meaning they contribute with a negligible
$0.4\,\ln(10)\,\sigma_\zeta$ to the luminosity error budget. Other
contributions amount to $0.4\,\ln(10)\,R_\zeta\,\sigma_{E(B-V)}$ for reddening,
$0.4\,\ln(10)\,\sigma_{BC}$ for BCs, and $2\sigma_\omega/\omega$ for parallaxes.
This implies that the uncertainty in BCs is dominant over the parallax error
when $\sigma_{BC}\gtrsim 2.2\,\sigma_\omega/\omega$. In other words,
when parallaxes are better than $0.5$\%, BCs are the dominant source of
uncertainty if $\sigma_{BC}\sim0.01$. This is shown by the purple line in
Figure \ref{fig:errors}, where the interplay among different uncertainties for a
target luminosity error is explored.
\vspace{-0.3cm}
\section{The colours of the Sun and Vega}\label{sec:colours}
\vspace{-0.1cm}
The solar colours provide an important benchmark point in many areas of
astronomy and astrophysics. The least model dependent, and arguably the best
method to determine them relies on solar twins \citep{m10,r12,c12}. Here we
use instead the formalism developed for \gaia synthetic magnitudes to compute
solar colors from a number of high fidelity, flux calibrated spectra. 
From the CALSPEC library we use a Kurucz model (sun\_mod\_001.fits) and a solar
reference spectrum (sun\_reference\_stis\_002.fits) which combines absolute
flux measurements from space and from the ground with a model spectrum
longward of 9600\AA\,\citep{colina96}. We also use the \cite{Thuillier04}
spectra for two solar active levels (about half of a solar cycle), where in
fact the difference between them is well below 1\,mmag across the \gaia filters
(hence we report only one set). Similarly, we can also use two spectra of
Vega available on the CALSPEC library to estimate its magnitudes and colours
(the Kurucz model alpha\_lyr\_mod\_002.fits, and alpha\_lyr\_stis\_008.fits
which intermingles a Kurucz model with HST-STIS measurement across part of the
\g and \bp bands). We remark that the \gaia system is tied to Vega
(assigned to have 0 magnitudes in all bands) using a slightly different Kurucz
model, and absolute flux calibration than CALSPEC. Hence, if we generate
magnitudes following the \gaia formalism, and believe CALSPEC to better match
the actual flux of Vega, it is not surprising that its magnitudes will be
slightly different from 0. 
As it can be seen from Table \ref{t:t1} there is excellent
agreement in the magnitudes and colours obtained from different spectral
templates, with differences of only a few mmag, comparable to the precision
reached by {\it Gaia}. At the level of $0.01$ mag, it is thus possible to
quote robust numbers for the Sun and Vega's magnitudes, independently of the adoped spectral
template, and flavour of zero-points and transmission curves. In the {\tt Vega}
system we have $G_\odot=-26.90$, corresponding to an absolute magnitude of
$M_{G,\odot}=4.67$, $(G_{\rm{BP}}-G)_\odot=0.33$,
$(G-G_{\rm{RP}})_\odot=0.49$ and $(G_{\rm{BP}}-G_{\rm{RP}})_\odot=0.82$ for the Sun,
and  $G=0.03$, $(G_{\rm{BP}}-G)=0.005$, $(G-G_{\rm{RP}})=0.01$ and
$(G_{\rm{BP}}-G_{\rm{RP}})=0.015$ for Vega. 
\begin{table*}
\centering
\caption{Predicted \gaia magnitudes and colours for the Sun and Vega in the
  {\tt Vega} and {\tt AB} systems. See Section \ref{sec:gaia} for a discussion of
  {\tt Gaia-pro}, {\tt Gaia-rev} and {\tt Gaia-DR2} realisations, and Section
  \ref{sec:colours} for a description of the spectral templates.}\label{t:t1}
\begin{tabular}{cccccccc}
\hline 
Object & template & \g & $G_{\rm{BP}}-G$ & $G-G_{\rm{RP}}$ & $G_{\rm{BP}}-G_{\rm{RP}}$ & system & realisation \\
\hline
Sun  & sun\_mod\_001.fits             & -26.792 &  0.257 & 0.241  &  0.498 & {\tt AB}   & {\tt Gaia-pro} \\
     &                                & -26.792 &  0.261 & 0.237  &  0.498 & {\tt AB}   & {\tt Gaia-rev} \\
     &                                & -26.897 &  0.333 & 0.490  &  0.823 & {\tt Vega} & {\tt Gaia-pro} \\
     &                                & -26.892 &  0.324 & 0.491  &  0.815 & {\tt Vega} & {\tt Gaia-rev} \\
     &                                & -26.895 &  0.329 & 0.488  &  0.818 & {\tt Vega} & {\tt Gaia-DR2} \\
     &                                &         &        &        &        &            &                \\  
     & sun\_reference\_stis\_002.fits & -26.792 &  0.257 & 0.242  &  0.499 & {\tt AB}   & {\tt Gaia-pro} \\
     &                                & -26.791 &  0.261 & 0.238  &  0.500 & {\tt AB}   & {\tt Gaia-rev} \\
     &                                & -26.897 &  0.333 & 0.491  &  0.824 & {\tt Vega} & {\tt Gaia-pro} \\
     &                                & -26.891 &  0.324 & 0.492  &  0.816 & {\tt Vega} & {\tt Gaia-rev} \\
     &                                & -26.894 &  0.330 & 0.489  &  0.819 & {\tt Vega} & {\tt Gaia-DR2} \\
     &                                &         &        &        &        &            &                \\
     &  Thuillier et al. (2004)       & -26.799 &  0.259 & 0.244  &  0.502 & {\tt AB}   & {\tt Gaia-pro} \\
     &                                & -26.798 &  0.263 & 0.240  &  0.503 & {\tt AB}   & {\tt Gaia-rev} \\
     &                                & -26.904 &  0.335 & 0.493  &  0.828 & {\tt Vega} & {\tt Gaia-pro} \\
     &                                & -26.898 &  0.326 & 0.493  &  0.819 & {\tt Vega} & {\tt Gaia-rev} \\
     &                                & -26.901 &  0.331 & 0.491  &  0.823 & {\tt Vega} & {\tt Gaia-DR2} \\
     &                                &         &        &        &        &            &                \\
Vega & alpha\_lyr\_mod\_002.fits      &  0.140  & -0.072 & -0.238 & -0.310 & {\tt AB}   & {\tt Gaia-pro} \\
     &                                &  0.134  & -0.057 & -0.243 & -0.300 & {\tt AB}   & {\tt Gaia-rev} \\
     &                                &  0.035  &  0.004 &  0.011 &  0.015 & {\tt Vega} & {\tt Gaia-pro} \\
     &                                &  0.034  &  0.006 &  0.010 &  0.016 & {\tt Vega} & {\tt Gaia-rev} \\
     &                                &  0.031  &  0.011 &  0.008 &  0.019 & {\tt Vega} & {\tt Gaia-DR2} \\
     &                                &         &        &        &        &            &                \\
     & alpha\_lyr\_stis\_008.fits     &  0.138  & -0.073 & -0.240 & -0.313 & {\tt AB}   & {\tt Gaia-pro} \\
     &                                &  0.132  & -0.058 & -0.246 & -0.304 & {\tt AB}   & {\tt Gaia-rev} \\
     &                                &  0.033  &  0.003 &  0.009 &  0.012 & {\tt Vega} & {\tt Gaia-pro} \\
     &                                &  0.032  &  0.005 &  0.008 &  0.012 & {\tt Vega} & {\tt Gaia-rev} \\
     &                                &  0.029  &  0.010 &  0.006 &  0.016 & {\tt Vega} & {\tt Gaia-DR2} \\
\hline
\end{tabular}
\end{table*}
Finally, in Table \ref{tab:tabler} we report extinction coefficients for the
\gaia filters, both average, and $\teff$-dependent ones. Users interested in
extinction coefficients at different values of temperature and/or metallicities
can easily derive them from our routines. 
\begin{table}
\centering
\caption{Extinction coefficients for \gaia filters. We report
  mean extinction coefficients $\langle R_\zeta\rangle$ and
  a linear fit valid for $5250 \le \teff \le 7000$\,{\rm{K}}.}\label{tab:tabler}
\smallskip
\begin{tabular}{lcccccc}
\hline
\hline 
\noalign{\smallskip}
 & & \multicolumn{4}{c}{$R_\zeta = a_0 + T_4\,(a_1 + a_2\,T_4) + a_3\,$[Fe/H]}\\
Filter & $\langle R_\zeta\rangle$ & \multispan4\hrulefill \\
 & & $a_0$ & $a_1$ & $a_2$ & $a_3$  \\
\noalign{\smallskip}
\hline
\noalign{\smallskip}
 $G$      & 2.740 & 1.4013 & 3.1406 & $-1.5626$ & $-0.0101$ \\
 $G_{BP}$ & 3.374 & 1.7895 & 4.2355 & $-2.7071$ & $-0.0253$ \\
 $G_{RP}$ & 2.035 & 1.8593 & 0.3985 & $-0.1771$ & \phantom{+}$0.0026$ \\
\noalign{\smallskip}
\hline
\noalign{\smallskip}
\end{tabular}
\begin{minipage}{0.5\textwidth}
  Based on the differences in the bolometric corrections for
  $E(B-V) = 0.0$ and $0.10$, assuming $\logg=4.1$,
  $-2.0 \le$ [Fe/H] $\le +0.25$,
  with [$\alpha$/Fe] = $-0.4, 0.0$ and $0.4$ at each [Fe/H]. Note
  that $T_4 = 10^{-4}\,T_{\rm eff}$. For a given nominal $E(B-V)$,
  the excess in any given $\zeta-\eta$ colour is
  $E(\zeta-\eta)=(R_\zeta-R_\eta)E(B-V)$, and the attenuation for a 
  magnitude $m_\zeta$ is $R_\zeta E(B-V)$.\\
\phantom{~~~~~~~~~~~~~~~~~~~~~~~}\\
\phantom{~~~~~~~~~~~~~~~~~~~~~~~}
\end{minipage}
\end{table}
\vspace{-0.8cm}
\section{Conclusions}
{\it Gaia}, not least its photometry, will induce a paradigm shift in many
areas of astronomy. However, to make full use of these data, colour
predictions from stellar fluxes are mandatory, as well as control of
systematics. We have expanded our previous investigations using MARCS stellar
fluxes to include \gaia \bpgrp
magnitudes. In doing so, we have explored the effects of implementing the two
different sets of bandpasses and zero-points that have been released with
\gaia DR2.  Differences are typically of few mmag only. Further, we have
generated a third set, which takes into a account the \cite{brown}
recommendations to provide the best match to observations. All of these
sets are available as part of our interpolation package for users to explore.
In examining the adopted zero-points,
we uncovered a magnitude-dependent offset in \gaia \g magnitudes. Albeit
small, this trend amounts to 30\,mmag over 10 magnitudes in the
\g band, which is larger than systematic effects at the 10\,mmag level
quoted by \cite{evans}.  This offset is relatively small, but it potentially
has a number of implications should \g magnitudes be used, e.g., to calibrate
distance indicators. Despite this offset, we regard \gaia magnitudes as an
incredible success, delivering magnitudes for a billion sources with an
accuracy within a few percent of CALSPEC. 

We also carried out an evaluation of the quality of our BCs, and their impact
on the luminosity error budget. \g and \rp magnitudes are typically better than
\bp in recovering bolometric fluxes, although averaging different bands is
probably advisable whenever possible.  Also, the systematic trend uncovered in
\g magnitudes does not impact bolometric fluxes too seriously, since
uncertainties in adopted stellar parameters and the performance of the
synthetic MARCS fluxes enter the error budget with a similar degree of
uncertainty. 
All our previous interpolation routines and scripts have now been updated to
include the \gaia system, and are available on GitHub (\href{https://github.com/casaluca/bolometric-corrections}{\tt github.com/casaluca/bolometric-corrections}).
A description of the files, and examples of their use can be found in
Appendix A of Paper II. 
\vspace{-0.8cm}
\section*{Acknowledgments}
We thank F.~De Angeli and P.~Montegriffo for useful correspondence,
and the referee G. Busso for the same kindness.
LC is the recipient of the ARC Future Fellowship FT160100402.
Parts of this research were conducted by the ARC Centre
of Excellence ASTRO 3D, through project number CE170100013. This work has made
use of data from the European Space Agency (ESA) mission {\it Gaia}
(\url{https://www.cosmos.esa.int/gaia}),
processed by the {\it Gaia} Data Processing and Analysis Consortium (DPAC,
\url{https://www.cosmos.esa.int/web/gaia/dpac/consortium}). Funding
for the DPAC has been provided by national institutions, in particular
the institutions participating in the {\it Gaia} Multilateral Agreement.
\vspace{-0.7cm}
\bibliographystyle{mn2e}
\bibliography{refs}

\begin{thebibliography}{19}
\providecommand{\natexlab}[1]{#1}

\bibitem[{{Altavilla} et~al.(2015)}]{altavilla}
{Altavilla} G. et~al., 2015, Astronomische Nachrichten, 336, 515

\bibitem[{{Andrae} et~al.(2018)}]{andrae}
{Andrae} R. et~al., 2018, arXiv:1804.09374

\bibitem[{{Bohlin}(2014)}]{bohlin14}
{Bohlin} R.~C., 2014, \aj, 147, 127

\bibitem[{{Cardelli} et~al.(1989){Cardelli}, {Clayton} \& {Mathis}}]{ccm89}
{Cardelli} J.~A., {Clayton} G.~C., {Mathis} J.~S., 1989, \apj, 345, 245

\bibitem[{{Carrasco} et~al.(2016)}]{carrasco}
{Carrasco} J.~M. et~al., 2016, \aap, 595, A7

\bibitem[{{Casagrande} \& {VandenBerg}(2014)}]{cv14}
{Casagrande} L., {VandenBerg} D.~A., 2014, \mnras, 444, 392

\bibitem[{{Casagrande} \& {VandenBerg}(2018)}]{cv18}
{Casagrande} L., {VandenBerg} D.~A., 2018, \mnras, 475, 5023

\bibitem[{{Casagrande} et~al.(2012){Casagrande}, {Ram{\'{\i}}rez},
  {Mel{\'e}ndez} \& {Asplund}}]{c12}
{Casagrande} L. et al., 2012, \apj, 761, 16

\bibitem[{{Colina} et~al.(1996){Colina}, {Bohlin} \& {Castelli}}]{colina96}
{Colina} L., {Bohlin} R.~C., {Castelli} F., 1996, \aj, 112, 307

\bibitem[{{Evans} et~al.(2018)}]{evans}
{Evans} D.~W. et~al., 2018, arXiv:1804.09368

\bibitem[{{Gaia Collaboration} et~al.(2018{\natexlab{a}}){Gaia Collaboration},
  {Brown}, {Vallenari}, {Prusti}, {de Bruijne}, {Babusiaux} \&
  {Bailer-Jones}}]{brown}
{Gaia Collaboration} et~al., 2018{\natexlab{a}}, arXiv:1804.09365
  
\bibitem[{{Gaia Collaboration} et~al.(2018{\natexlab{b}})}]{babu}
{Gaia Collaboration} et~al., 2018{\natexlab{b}}, arXiv:1804.09378

\bibitem[{{Gustafsson} et~al.(2008){Gustafsson}, {Edvardsson}, {Eriksson},
  {J{\o}rgensen}, {Nordlund} \& {Plez}}]{g08}
{Gustafsson} B. et al., 2008, \aap, 486, 951

\bibitem[{{Jordi} et~al.(2010)}]{jordi10}
{Jordi} C. et~al., 2010, \aap, 523, A48

\bibitem[{{Mel{\'e}ndez} et~al.(2010){Mel{\'e}ndez}, {Schuster}, {Silva},
  {Ram{\'{\i}}rez}, {Casagrande} \& {Coelho}}]{m10}
{Mel{\'e}ndez} J. et al., 2010, \aap, 522, A98+

\bibitem[{{Pancino} et~al.(2012)}]{pancino}
{Pancino} E. et~al., 2012, \mnras, 426, 1767

\bibitem[{{Ram{\'{\i}}rez} et~al.(2012)}]{r12}
{Ram{\'{\i}}rez} I. et~al., 2012, \apj, 752, 5

\bibitem[{{Riello} et~al.(2018)}]{riello18}
{Riello} M. et~al., 2018, arXiv:1804.09367

\bibitem[{{Thuillier} et~al.(2004){Thuillier}, {Floyd}, {Woods}, {Cebula},
  {Hilsenrath}, {Hers{\'e}} \& {Labs}}]{Thuillier04}
{Thuillier} G. et al., 2004, Advances in Space Research, 34, 256

\end{thebibliography}

\end{document}